\documentclass[prd,twocolumn,reprint,preprintnumbers,nofootinbib,superscriptaddress,longbibliography]{revtex4-1} 
\usepackage{graphicx}
\usepackage{dcolumn}
\usepackage[utf8]{inputenc}
\usepackage{bm}
\usepackage[caption=false]{subfig}
\usepackage{amssymb}
\usepackage{float}
\usepackage{hyperref}
\usepackage[T1]{fontenc} 
\usepackage{subfig}
\usepackage{dsfont}
\usepackage{slashed}
\usepackage{color}
\usepackage{amsmath}
\usepackage{mathtools}
\usepackage[section]{placeins}
\usepackage{braket}
\usepackage{upgreek}
\usepackage[bottom]{footmisc}
\usepackage[normalem]{ulem}
\usepackage{lipsum}
\usepackage[utf8]{inputenc}
\usepackage{booktabs}

\begin{document}

\title{High-Energy Nuclear Recoils from Boosted Dark Matter for the LZ 248-keV Event:\\
Beyond the Halo-Dependent High-Velocity Tail
} 

\author{Haider Alhazmi}
\email{hmalhazmi@jazanu.edu.sa}
\affiliation{Department of Physical Sciences, Jazan University, Jazan 45142, Saudi Arabia}

\author{Doojin Kim}
\email{doojin.kim@usd.edu}
\affiliation{Department of Physics, University of South Dakota, Vermillion, SD 57069, USA}

\author{Kyoungchul Kong}
\email{kckong@ku.edu}
\affiliation{Department of Physics and Astronomy, University of Kansas, Lawrence, KS 66045, USA}

\author{Jong-Chul Park}
\email{jcpark@cnu.ac.kr}
\affiliation{Department of Physics and Institute for Sciences of the Universe, Chungnam National University, Daejeon 34134, Korea}

\author{Seodong Shin}
\email{sshin@jbnu.ac.kr}
\affiliation{Laboratory of Symmetry and Structure of the Universe, Department of Physics, Jeonbuk National University, Jeonju, Jeonbuk 54896, Korea}
\affiliation{School of Physics, Korea Institute for Advanced Study, Seoul 02455, Korea}
\affiliation{Particle Theory and Cosmology Group, Center for Theoretical Physics of the Universe, Institute for Basic Science, Daejeon 34126, Korea
}

\begin{abstract}
The LUX-ZEPLIN (LZ) Collaboration has reported a nuclear-recoil candidate at \(E_R=248\pm23_{\rm stat}\pm23_{\rm sys}\)~keV, with a maximum local significance of \(3.4\sigma\) and a global significance of \(2.6\sigma\). A prominent interpretation invokes heavy halo dark matter near an inelastic threshold and therefore depends sensitively on the poorly constrained high-speed tail of the Galactic velocity distribution. In this Letter, we propose a qualitatively different possibility based on light boosted dark matter (BDM), whose incident energy is determined primarily by the dark-sector mass spectrum. We consider multi-component scenarios in which the boosted state scatters elastically or inelastically off xenon nuclei. For elastic scattering, pseudoscalar-mediated momentum dependence suppresses low-energy recoils. Near-threshold endothermic scattering of a nearly monochromatic BDM flux can instead confine the signal between kinematically determined recoil endpoints, suppressing events in both the low- and high-energy sidebands. The upscattered state may furthermore decay invisibly within the dark sector, preserving a single-nuclear-recoil signature without requiring it to be detector-stable. We present representative benchmark spectra and discuss complementary tests using other target nuclei and large-volume liquid-scintillator experiments.
\end{abstract}

\maketitle

\noindent {\bf Introduction.} 
The LUX-ZEPLIN (LZ) Collaboration has analyzed the nuclear-recoil data corresponding to a 2.84 tonne-years exposure up to recoil energy $E_R\simeq 269.9$~keV, and has reported one excessive event at $E_R=248\pm23_{\rm stat}\pm23_{\rm sys}$~keV, with a global background-only significance of 2.6$\sigma$~\cite{LZ:2026axp}. 
If interpreted as dark matter (DM), the crucial requirement is not only to accommodate such a high-energy nuclear-recoil event but also to avoid predicting a more conspicuous population below $\sim 200$~keV. 

The community quickly responded by proposing a range of scenarios capable of satisfying the phenomenological requirements~\cite{Fan:2026kxx, Freese:2026sga, Wu:2026nhi, Lou:2026idn, Yin:2026jnn, DiMauro:2026ldr, Visinelli:2026kgt, Yamashita:2026ump, Chattopadhyay:2026ryw, Smirnov:2026aqk, Du:2026guj, Rodd:2026tyn, McCabe:2026crm, Jeesun:2026vzo, Unwin:2026rdp}. 
Many of the interpretations invoke upscattering of virialized Galactic-halo DM, typically with a mass of several hundred GeV to the TeV scale and a mass splitting of $\mathcal O$(100 keV). 
While this mechanism naturally shifts the recoil spectrum toward higher energies, the production of $\sim 248$~keV events generally relies on the sparsely populated and astrophysically uncertain high-speed tail part of the halo distribution. 

This observation motivates an alternative possibility where the required incident energy is supplied by boosted dark matter (BDM)~\cite{Belanger:2011ww,Agashe:2014yua,Kong:2014mia,Kim:2016zjx,Giudice:2017zke,Alhazmi:2016qcs,Kim:2018veo}, without necessarily demanding either a TeV-scale DM, a velocity near the halo cutoff, or both. 
Furthermore, in a multi-component realization, a nearly monochromatic flux and near-threshold endothermic scattering can confine the recoil spectrum between controlled endpoints, while an invisibly decaying upscattered state leaves only a single nuclear recoil without requiring a detector-scale lifetime.
We therefore investigate which BDM interactions can accommodate the high-energy recoil event observed by LZ, while sufficiently suppressing lower-energy events. 
We also point out that solar neutrino experiments such as Borexino and JUNO may be able to test the DM explanations of the LZ event.

\medskip
\noindent {\bf BDM Interpretations and Benchmarks.}
A high-energy nuclear-recoil event without a corresponding lower-energy excess requires both an energetic incident flux localized in a rather narrow regime and a mechanism that preferentially populates the high-recoil region.  
Canonical multi-component BDM scenarios~\cite{Belanger:2011ww,Agashe:2014yua,Kong:2014mia,Kim:2016zjx} naturally provide the former via annihilation or decay of a heavier DM component into a lighter one.
Along this line, we focus on two such representative scenarios: elastically scattering BDM (eBDM)~\cite{Belanger:2011ww,Agashe:2014yua,Kong:2014mia} and inelastically scattering BDM ($i$BDM)~\cite{Kim:2016zjx,Giudice:2017zke,Chatterjee:2018mej,Heurtier:2019rkz,Kim:2020ipj,DeRoeck:2020ntj}. 

In general, the boosted flux alone does not guarantee the required spectral shape. 
For example, conventional elastic scattering generally populates the low-energy recoil region as well. 
Thus, a viable eBDM explanation requires dynamics that strongly suppresses scattering at small momentum transfer (denoted by $q$).
We therefore consider eBDM interacting through a pseudoscalar mediator as an illustrative benchmark~\cite{Alhazmi:2020fju}. 
For $i$BDM, by contrast, the localization of the recoil spectrum can also be realized by the kinematics of the endothermic process:
\begin{align}
    \chi_1 + N \to \chi_2 + N\,,
    \label{eq:BDMscattering}
\end{align}
where $\chi_1$ is the incoming light BDM particle scattering with a target nucleus $N$ and $\chi_2$ is a heavier dark-sector state produced via this inelastic scattering, whose threshold can suppress or completely eliminate low-energy nuclear recoils. 
The $i$BDM interpretation consequently depends less on the mediator choice at the level of recoil localization, although the detailed spectral shape and normalization remain interaction-dependent. 
We thus consider a dark photon $A'$ and a pseudoscalar $a$ as representative $i$BDM mediators.
For the BDM spin, we consider Dirac or pseudo-Dirac fermions, as well as complex scalars~\cite{Giudice:2017zke}.\footnote{In the inelastic scalar realization~\cite{Giudice:2017zke}, the complex scalar is decomposed into two nondegenerate real scalar mass eigenstates.}

In multi-component eBDM scenarios~\cite{Belanger:2011ww,Agashe:2014yua,Kong:2014mia,Alhazmi:2016qcs,Kim:2018veo}, annihilation of the lighter component into Standard-Model particles may also play an important role in determining its relic abundance~\cite{Belanger:2011ww} and can lead to significant late-time energy injection. 
To evade cosmological and astrophysical constraints, such an annihilation is preferably $p$-wave dominated or otherwise suppressed at late times~\cite{Kamada:2021muh,Kim:2023onk,Kim:2024ltz}. 
Nevertheless, we also include Dirac-fermion eBDM interacting through a dark photon, as this is one of the most widely studied BDM benchmarks, assuming that additional ingredients beyond those considered in Refs.~\cite{Kamada:2021muh,Kim:2023onk,Kim:2024ltz} resolve the associated cosmological constraints. 
A dedicated study of the cosmological viability of the $i$BDM scenarios---including their thermal histories and late-time annihilation rates---is deferred to future work. 

We note that the energy-localizing feature is generally lost for BDM sources with a broad incident-energy spectrum. 
After integrating over such a continuum, higher-energy components open progressively wider recoil intervals and populate recoil energies well below the observed event; this tendency is often reinforced by fluxes and scattering cross sections weighted toward lower energies. 
Continuum BDM scenarios (e.g., Refs.~\cite{Bringmann:2018cvk, Ema:2018bih, Cappiello:2019qsw, Dent:2019krz, Jho:2020sku, Cho:2020mnc, Jho:2021rmn, Das:2021lcr, Guha:2024mjr, Guha:2025tjf, Chao:2021orr, DeRocco:2019jti, Bhalla:2025vnq}) are therefore generically disfavored by the absence of a lower-energy excess unless an additional threshold, sharp spectral feature, or sufficiently strong momentum dependence suppresses that population.

\medskip
\noindent {\bf BDM Kinematics.} 
We now consider the (inelastic) scattering of a BDM particle $\chi_1$ of mass $m_1$ with fixed incident energy $E_1$ on a (stationary) target nucleus $N$ of mass $m_N$, producing $\chi_2$ with mass $m_2$ as in Eq.~(\ref{eq:BDMscattering}). 
In multi-component annihilating BDM scenarios, $E_1$ is identified as the mass of the heavier DM component that is the dominant relic in the present universe, and hence the distribution of $E_1$ is given by a Dirac delta function.
Elastic scattering is straightforwardly recovered by setting $m_2=m_1$. 
For this process, the center-of-mass energy $\sqrt{s} = \sqrt{m_1^2+m_N^2+2m_N E_1}$. 
The kinematically allowed recoil energy spans $E_R^-$ to $E_R^+$:
\begin{equation}
E_R^\pm=\frac{2E_1^\ast E_2^\ast
\pm 2p_1^\ast p_2^\ast-m_1^2-m_2^2}{2m_N}\,,
\end{equation}
where center-of-mass quantities $E_{1,2}^*$ and $p_{1,2}^*$ are
\begin{equation}
E_{1,2}^\ast=
\frac{s+m_{1,2}^2-m_N^2}{2\sqrt{s}}\,,
\quad
p_{1,2}^\ast=
\frac{\lambda^{1/2}(s,m_{1,2}^2,m_N^2)}{2\sqrt{s}}\,,
\end{equation}
with $\lambda$ defined as \(\lambda(x,y,z)=(x-y-z)^2-4yz\).

For an endothermic transition, the channel opens when $E_1 \geq E_{1, {\rm th}}=m_2+\frac{m_2^2-m_1^2}{2m_N}$. 
At the threshold, the recoil energy is uniquely fixed to
\begin{equation}
    E_{R,{\rm th}}=\frac{m_2^2-m_1^2}{2(m_N+m_2)}\,. \label{eq:ErecoTh}
\end{equation}
Thus, we find that for $m_2 \ll m_{\rm Xe}$, localizing the xenon recoil near the LZ excess event requires
\begin{equation}
m_2^2-m_1^2 \simeq2m_{\rm Xe}E_R \simeq(246~\mathrm{MeV})^2
\left(\frac{E_R}{248~\mathrm{keV}}\right).
\end{equation}
For example, a nearly monoenergetic BDM flux with $E_1$ slightly above $E_{1,{\rm th}}({\rm Xe})$ therefore produces a narrow window around 248~keV without relying on the high-speed tail of the Galactic halo. 
In contrast to heavy-DM interpretations such as the thermal Higgsino scenario~\cite{Rodd:2026tyn}, the benchmarks considered below have upper recoil endpoints well below 350~keV, leaving no truth-level events---and only negligible detector-smeared leakage---in the LZ high-energy sideband at the current exposure.
Furthermore, since the threshold in Eq.~\eqref{eq:ErecoTh} decreases with $m_N$, there exists a target-selective region:
\begin{equation}
E_{1,\mathrm{th}}(\mathrm{Xe}) <E_1 <E_{1,\mathrm{th}}(\mathrm{C}) <E_{1,\mathrm{th}}(p)\,,
\label{eq:wedge}
\end{equation}
for which the transition is allowed in xenon but kinematically forbidden on the carbon and hydrogen targets relevant to Borexino and JUNO. 

The differential cross section with respect to $E_R$ is given by
\begin{equation}
 \frac{d\sigma}{dE_R}
 =
 \frac{m_N}
 {8\pi\lambda(s,m_1^2,m_N^2)}
 \overline{\left|
 \mathcal M_N(E_1,E_R)
 \right|^2}\,,
 \label{eq:generic-cross-section}
\end{equation}
where the bar denotes the usual spin-averaged sum. 
The amplitude $\mathcal M_N$ incorporates both the underlying particle-level interaction and the corresponding nuclear matrix elements. 
Its momentum-transfer dependence may be organized schematically in terms of nuclear response functions $\mathcal R_N(Q^2)$ with $Q^2\equiv-q^2= 2m_N E_R$, whose form depends on the Lorentz and isospin structure of the interaction.

For a vector mediator coupled coherently to the nuclear charge, the dominant response is
\begin{equation}
\mathcal R_N^V(Q^2)
=Z^2F_{\rm H}^2(Q^2)\,,
\label{eq}
\end{equation}
where $F_{\rm H}$ is the Helm form factor~\cite{Helm:1956zz,Lewin:1995rx}. 
For a pseudoscalar mediator, the nuclear response cannot be described by the Helm form factor. Instead, the recoil spectrum can be expressed schematically as
\begin{equation}
    \frac{d\sigma_N}{dE_R} \propto \frac{{\cal K}_\chi(Q^2)}
     {(Q^2+m_a^2)^2} \sum_{\tau,\tau'=0,1} c_\tau(Q^2)c_{\tau'}(Q^2) W_{\Sigma''}^{\tau\tau'}(Q), 
\end{equation}
where $W_{\Sigma''}^{\tau\tau'}$ is the isotope-dependent spin-longitudinal nuclear response and $c_\tau$ denotes the isoscalar/isovector nucleon coupling~\cite{Fitzpatrick:2012ix, Anand:2013yka}. 
The $Q^2$-dependent kernel is given by ${\cal K}_\chi=Q^4~(Q^2)$ for the fermionic (scalar) DM interaction $i g_{12}a\bar{\chi}_2\gamma^5\chi_1$ ($\mu_{12}a \chi_2^\dag\chi_1$).
As mentioned earlier, the momentum dependence may dynamically suppress low-energy recoils, whereas an endothermic transition removes them kinematically through a positive lower recoil endpoint. 
We next investigate benchmark interactions, identifying parameter choices that produce the LZ high-energy event without an excessive lower-energy population.

\medskip
\noindent {\bf Parameter Choices and Results.}
The practical parameter ranges for our initial scan are
\begin{eqnarray}
    &&m_1\in[5,200]~{\rm MeV},\qquad m_2\in[230,320]~{\rm MeV},\nonumber \\
    &&E_1-m_2\in [0.2,2.5]~{\rm MeV}.
\end{eqnarray}
We further require that the accepted parameter points satisfy the following requirements:
\begin{align} 248~\mathrm{keV}\in[E_R^-,E_R^+], \qquad E_R^-\gtrsim(180\text{--}200)~\mathrm{keV}\,, 
\end{align}
together with the xenon-only kinematic condition given in Eq.~\eqref{eq:wedge}.
The precise lower bound imposed on $E_R^-$ depends on the adopted event selection. 
Representative benchmark points that we identified are listed in Table~\ref{tab:spectralbenchmarks}.
The relevant model details and differential cross sections are described in e.g., Refs.~\cite{Kim:2016zjx,Giudice:2017zke,Kim:2020ipj,DeRoeck:2020ntj,Alhazmi:2020fju,Alhazmi:2025nvt}.
For illustration, we set \(m_1=10~\mathrm{MeV}\) and \(m_{A',a}=0.2~\mathrm{GeV}\) for all benchmark points.
The four eBDM benchmarks share the same masses and incident energy, while their spectral shapes differ because of the DM spin and mediator-dependent matrix elements. 
For the $i$BDM scenarios, since $m_2>m_1+m_{A',a}$, the upscattered state $\chi_2$ can decay invisibly through an on-shell mediator, i.e., $\chi_2\to \chi_1+(A'/a \to 2\chi_1)$.\footnote{Denoting the transition, $\chi_1$ diagonal, and SM-mediator couplings by \(g_{12}\), \(g_{11}\), and \(g_{\rm SM}\), respectively, we assume \(g_{12}\gg g_{11}\gg g_{\rm SM}\), so that the transition interaction dominates while the on-shell mediator decays predominantly into \(\chi_1\chi_1\).} 
The LZ signature therefore remains a single nuclear recoil even if the decay occurs promptly inside the detector, eliminating the detector-scale lifetime requirement encountered in visibly decaying inelastic ambient DM scenarios.
\begin{table}[t]
    \centering
    \begin{tabular}{lccccc}
    \toprule
     Label & BDM types & $m_1$ & $m_2$ & $E_1$ & Xe $E_R$ interval \\
     \midrule
     D-iV & Dirac $i$BDM & 10 & 247.004 & 247.263 & [244.1, 253.0] keV \\
     D-iP & Dirac $i$BDM & 10 & 246.722 & 246.980 & [243.5, 252.5] keV \\
     S-iV & Scalar $i$BDM & 10 & 246.997 & 247.256 & [244.1, 253.0] keV \\    
     S-iP & Scalar $i$BDM & 10 & 245.220 & 246.730 & [199.9, 300.1] keV \\
      & eBDM & 10 & --- & 135.962 & [0, 300] keV \\
    \bottomrule    
    \end{tabular}
    \caption{Benchmark points used for the spectral comparison. 
    All masses and energies not otherwise specified are in MeV. 
    $m_{A',a}$ are set to be 0.2 GeV. 
    In the label, ``V'' and ``P'' imply dark-photon and pseudoscalar mediations, respectively. 
    The $i$BDM benchmark points are 10 keV above their exact xenon thresholds and are tuned so that the efficiency-weighted narrow spectrum is centered at 248 keV. 
    For eBDM, the four combinations, D/S-eV/P, are implicitly assumed. 
    }
    \label{tab:spectralbenchmarks}
\end{table}

\begin{figure*}[t]
    \centering    
    \includegraphics[width=0.9\linewidth]{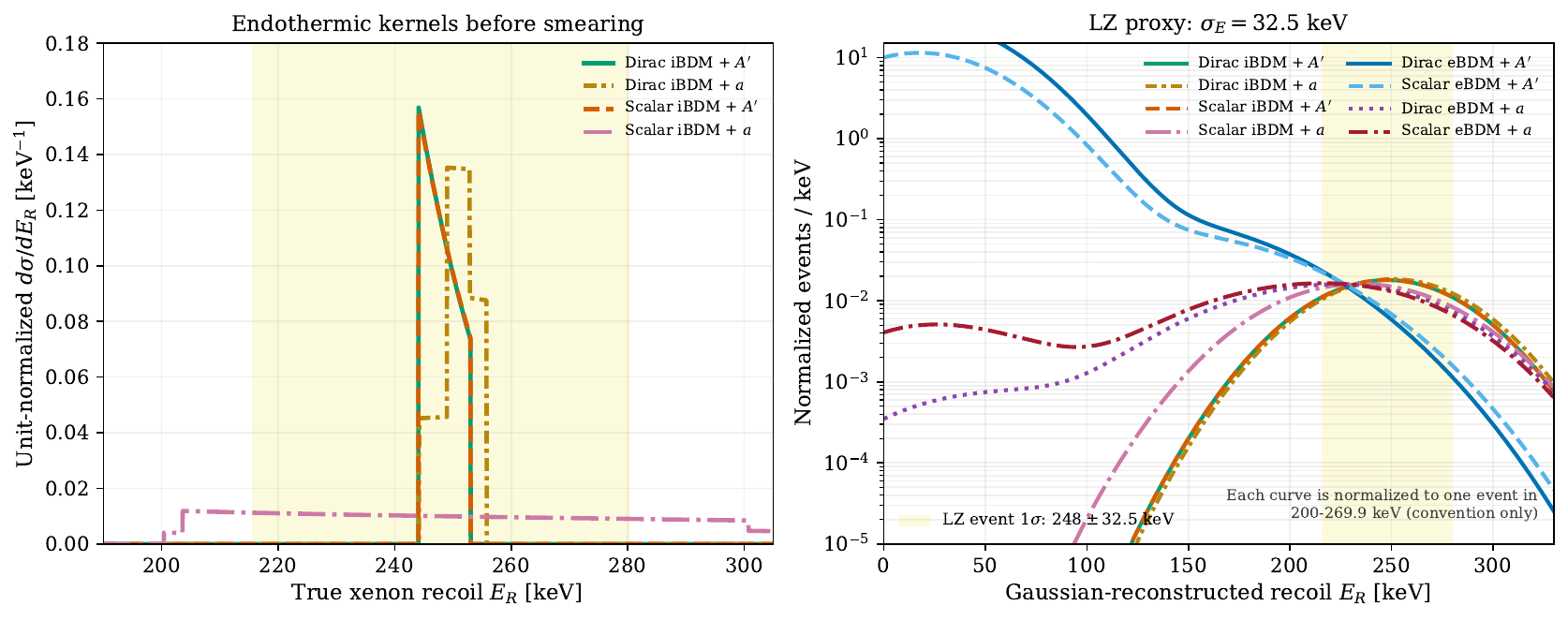}
    \caption{Recoil energy spectra for the benchmarks in Table~\ref{tab:spectralbenchmarks}.  
    Left: unit-normalized true $i$BDM spectra.
    Right: spectra after the efficiency proxy and Gaussian smearing, each normalized to one reconstructed event between 200 and 269.9 keV. 
    The $i$BDM spectra provide illustrative efficiency-folded spectra, whereas elastic dark-photon scattering produces an enormous low-energy population. 
    We take $m_{A',a}=0.2$~GeV. 
    The smooth LZ efficiency proxy has 50\% points at 5.4 and 269.9 keV and a 96\% plateau.  
    The yellow band shows only the event's combined $1\sigma$ interval, $248\pm32.5$ keV.
    }    
    \label{fig:spectra}
\end{figure*}

Figure~\ref{fig:spectra} displays the recoil spectra corresponding to the benchmarks in Table~\ref{tab:spectralbenchmarks}.
The left panel presents the true recoil spectra for the four $i$BDM benchmarks before detector effects are applied, with each spectrum normalized to the unit area. 
At this level, the differences caused by the DM spin and mediator-dependent matrix elements are visible. 
Nevertheless, all four spectra are sharply localized around the desired $E_R$ value by the endothermic kinematics.

In the right panel, we weight the true spectra by a smooth analytic proxy for the LZ signal efficiency and convolve them with an illustrative Gaussian response. 
The efficiency proxy is anchored to the reported 50\% efficiency points of the central black curve in Fig.~S2 of Ref.~\cite{LZ:2026axp}; it should not be interpreted as an official LZ detector-response model, though. 
For comparison of the spectral shapes, each efficiency-weighted and smeared spectrum is normalized to one reconstructed event in the high-energy interval \(E_R\in[200,269.9]~{\rm keV}\). 
The 200-keV boundary separates the high-energy normalization region containing the candidate from the lower-energy region used to compare the accompanying recoil populations.
The upper boundary corresponds to the recoil energy at which the reported LZ signal efficiency falls to \(50\%\)~\cite{LZ:2026axp}. 
This interval is adopted only as a common normalization convention and should not be interpreted as either an official LZ recoil-energy window or the \(1\sigma\) uncertainty interval of the observed event.
The yellow band indicates the combined $1\sigma$ interval of the candidate, 
\begin{align} 
E_R=248\pm \sqrt{23^2+23^2} =248\pm32.5~\mathrm{keV}\,, \end{align}
where the statistical and systematic uncertainties have been combined in quadrature.
For this exploratory comparison, we also use $\sigma_E=32.5~\mathrm{keV}$ as the Gaussian-smearing width. 
This choice is illustrative and again should not be regarded as the official LZ energy-resolution function.

After the efficiency weighting and Gaussian smearing, the reconstructed spectra of all four $i$BDM benchmarks peak near $248$ keV and show similar shapes. 
This demonstrates that the localization of the signal is primarily controlled by the endothermic kinematics rather than by the detailed mediator structure.
By contrast, the momentum-suppressed pseudoscalar-mediated eBDM scenarios (red-dotdashed and purple-dotted in the right panel of Fig.~\ref{fig:spectra}) produce broader spectra.
They suppress the low-energy recoil population relative to the dark-photon-mediated eBDM scenarios (blue-solid and cyan-dashed) and can retain appreciable support near the observed event. 
However, low-energy recoils are not kinematically forbidden in these elastic scenarios, making the explanation less automatic than in the endothermic $i$BDM case. 
For comparison, the dark-photon-mediated eBDM benchmarks predict much larger recoil populations below $200$ keV when normalized to one event in the LZ high-energy window. 
They are therefore strongly disfavored as explanations of an isolated $248$ keV event.

We note that the normalization required by the LZ event can be expressed independently of the BDM production mechanism as \(\mathcal F_1 \sigma \simeq 2.4\times10^{-36}~{\rm s}^{-1}\) with $\mathcal F_1$ being the $\chi_1$ flux. 
Representative mediator benchmarks consistent with existing constraints (e.g., kinetic mixing parameter $\epsilon\sim 10^{-3}$ at $m_{A'}=0.2$~GeV for the dark-photon scenario) typically yield $\sigma\lesssim10^{-37}~{\rm cm^2}$, implying a required flux of $\mathcal{O}(10)~{\rm cm^{-2}s^{-1}}$. 
This exceeds the canonical Galactic annihilation flux~\cite{Agashe:2014yua} for our benchmark masses by three-to-four orders of magnitude. 
Thus, while the endothermic $i$BDM framework naturally accounts for the recoil morphology, its absolute normalization calls for an enhanced or localized source. 
Our results should therefore be interpreted as a source-independent demonstration of the recoil morphology attainable with BDM, together with a quantitative normalization target that any complete production scenario must satisfy. 
In this sense, the framework trades the sensitivity of ambient inelastic DM to the extreme Galactic velocity tail for the requirement of an enhanced or localized boosted flux.
Possible mechanisms that may address this flux challenge include Sommerfeld-enhanced~\cite{Arkani-Hamed:2008hhe} or resonant present-day annihilation~\cite{Ibe:2008ye}, a steep DM density spike~\cite{Gondolo:1999ef, Fields:2014pia, Kim:2017qaw}, decays of a long-lived parent population~\cite{Bhattacharya:2014yha}, or a nearby source such as captured DM in the Sun~\cite{Berger:2014sqa, Kong:2014mia}. 
A complete realization requires a dedicated treatment of the associated cosmological, astrophysical, and propagation constraints, which we leave for future work.

\begin{figure}[h]
    \centering
    \includegraphics[width=\linewidth]{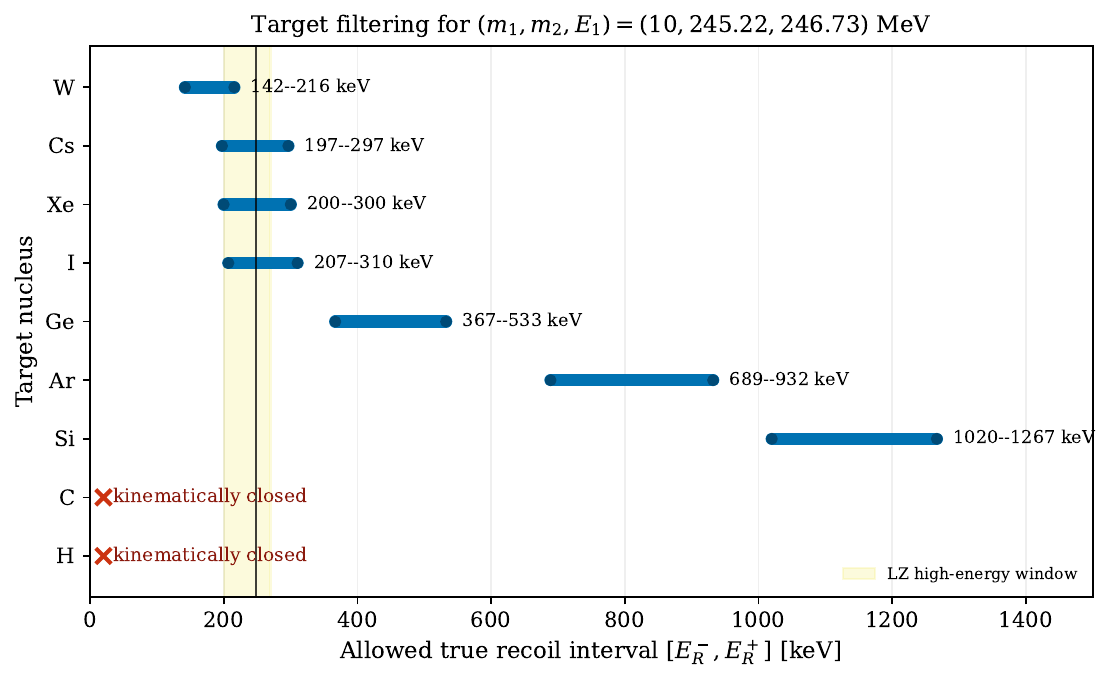}
    \caption{Kinematic recoil intervals for the baseline endothermic benchmark. 
    These intervals do not include nuclear response or detector thresholds. 
    Carbon and hydrogen are closed, while sufficiently heavy targets are open.  
    A spin-dependent pseudoscalar operator can nevertheless suppress spin-zero nuclei such as $^{40}$Ar; only kinematics is displayed.
    }
    \label{fig:targets}
\end{figure}

\medskip
\noindent 
{\bf Complementary Experimental Tests.}
An attractive feature of the BDM interpretations is that they can be tested independently by other xenon-based DM direct-detection experiments, notably XENONnT~\cite{XENON:2024wpa} and PandaX-4T~\cite{PandaX:2018wtu}. 
Complementary tests may also be performed using experiments with argon, cesium, iodine, tungsten, germanium, or silicon targets. 
Figure~\ref{fig:targets} shows the kinematically allowed recoil-energy interval for each target nucleus for our baseline endothermic benchmark, S-iP:~\((m_1,\,m_2,\,E_1)=(10,\,245.220,\,246.730)\)~MeV, without including nuclear-response functions, detector thresholds, or selection efficiencies. 
This benchmark predicts recoil energies of approximately $367$--$533~\mathrm{keV}$ in germanium, $689$--$932~\mathrm{keV}$ in argon, and 1,020--1,267~keV in silicon. 
Heavier target nuclei, such as xenon, iodine, cesium, and tungsten, therefore provide the most direct kinematic cross-checks of the LZ event.

The relative sensitivities of these experiments, however, depend strongly on the interaction operator. 
For a coherently enhanced off-diagonal vector interaction, an argon experiment could provide a powerful complementary test, provided that its analysis covers the predicted high-recoil-energy interval. 
By contrast, a pseudoscalar interaction probes the longitudinal nuclear-spin response and preferentially selects isotopes with nonzero nuclear spin. 
In that case, independent xenon measurements may provide the cleanest near-term cross-check, although iodine- and cesium-based experiments could also offer valuable complementary sensitivity.

Complementary searches can also be conducted with large-volume liquid-scintillator-based neutrino experiments with low visible-energy thresholds, such as Borexino~\cite{Borexino:2008gab,Borexino:2012udf} and JUNO~\cite{JUNO:2021vlw}. 
The published Borexino Phase-II analysis covers recoil-electron kinetic energies from $0.19$ to $2.93~\mathrm{MeV}$~\cite{Borexino:2017rsf}.
These data can constrain our eBDM scenarios if the mediator couples appreciably to electrons and the resulting electron-recoil spectrum overlaps this energy range. 
JUNO can provide a particularly powerful complementary probe because of its much larger liquid-scintillator target mass. 
If a dedicated low-energy analysis achieves a visible-energy threshold of $E_{\rm vis}\lesssim 200$ keV, as envisaged with its low-energy trigger, JUNO could strongly test eBDM coupled to nucleons through scattering on its enormous number of free-proton targets~\cite{Choi:2025wbw}. 
Proton quenching must, however, be included when converting the proton recoil energy $T_p$ into the observable visible energy.
On the other hand, for our optimized endothermic benchmarks, the incident $\chi_1$ energy lies below the thresholds for both \(\chi_1 p \to \chi_2 p\) and \(\chi_1{}^{12}\mathrm{C}\to\chi_2{}^{12}\mathrm{C}\), so the corresponding in-detector upscattering signals are kinematically forbidden at JUNO. 

\medskip
\noindent {\bf Conclusions.}
The isolated high-energy nuclear-recoil event reported by LZ is compatible with a DM origin, although a single event is insufficient to establish such an interpretation. 
In this Letter, we have identified a qualitatively different interpretation based on light BDM, whose incident energy is determined primarily by the dark-sector mass spectrum, rather than on heavy halo DM drawn from the poorly constrained high-speed tail of the Galactic velocity distribution.

Multi-component BDM scenarios provide a natural framework for realizing this possibility. 
In eBDM, a pseudoscalar mediator introduces momentum-transfer suppression that enhances the relative importance of high-energy recoils and substantially reduces the low-energy event rate compared with vector-mediated elastic scattering. 
In endothermic $i$BDM, by contrast, the lower recoil endpoint can be shifted above the low-energy search region, thereby kinematically eliminating low-energy recoils.
Combined with a nearly monochromatic incident flux, the same kinematics can also place the upper endpoint below the high-energy sideband, while an invisible decay such as \(\chi_2\to3\chi_1\) preserves a single-nuclear-recoil signature without requiring \(\chi_2\) to be detector-stable.
Among the scenarios considered here, optimized endothermic $i$BDM therefore provides the cleanest explanation of an isolated event near $248$ keV, whereas elastic pseudoscalar-mediated BDM remains a viable but more nuclear-response-dependent possibility.

The BDM interpretation can be tested not only by independent xenon-based experiments, such as XENONnT and PandaX-4T, but also by experiments employing argon, cesium, iodine, tungsten, germanium, or silicon targets. 
Measurements with different target nuclei would probe both the characteristic endothermic kinematics and the operator-dependent nuclear response, thereby providing an important means of discriminating among the proposed explanations.

Large-volume neutrino experiments offer further complementary opportunities. 
A dedicated reanalysis of the Borexino electron-recoil data could constrain scenarios with appreciable couplings to electrons. 
JUNO, with its enormous number of free-proton and carbon targets, could strongly test eBDM coupled to nucleons if a sufficiently low visible-energy threshold is achieved. 
Conversely, our optimized endothermic benchmarks predict no direct $\chi_1 p\to\chi_2 p$ or $\chi_1{}^{12}\mathrm{C}\to\chi_2{}^{12}\mathrm{C}$ events because both channels are kinematically closed.
Dedicated low-threshold searches at JUNO and complementary analyses of existing Borexino data would therefore provide valuable and qualitatively distinct tests of the DM interpretations of the LZ event.

\medskip
\acknowledgements
The work of DK is supported in part by the National Science Foundation (NSF) through Grant No. PHY-2609913.
The work of KK is supported in part by the NSF through Grant No. PHY-2609759. 
The work of JCP is supported by the National Research Foundation of Korea (NRF) grant funded by the Ministry of Science and ICT (RS-2024-00356960) and by the Global - Learning \& Academic research institution for Master's/PhD students and Postdocs (G-LAMP) Program of the NRF grant funded by the Ministry of Education (RS-2025-25442707). 
The work of SS is supported by the NRF with Grant No. RS-2025-00562917 and partly by the IBS fund IBS-R018-D1.

\bibliography{ref}

\end{document}